\documentclass[sigconf]{acmart}

\AtBeginDocument{%
  }

\usepackage{color,soul}
\usepackage{subcaption}
\usepackage{algpseudocode}
\usepackage{algorithm}
\usepackage[framemethod=tikz]{mdframed}

\begin{document}

\title{Consider What Humans Consider: Optimizing Commit Message Leveraging Contexts Considered By Human}

\author{Jiawei Li}
\orcid{0000-0002-4434-4812}
\affiliation{%
  \institution{University of California, Irvine}
  \city{Irvine}
  \country{USA}
}
\email{jiawl28@uci.edu}

\author{David Faragó}
\orcid{0009-0006-2380-6076}
\affiliation{%
  \institution{Innoopract, QPR Technologies}
  \city{}
  \country{Germany}
}
\email{drdavidfarago@gmail.com}

\author{Christian Petrov}
\orcid{0000-0001-8776-4289}
\affiliation{%
  \institution{Innoopract, Mediform}
  \city{}
  \country{Germany}
}
\email{christian.petrov@mediform.io}

\author{Iftekhar Ahmed}
\orcid{0000-0001-8221-5352}
\affiliation{%
  \institution{University of California, Irvine}
  \city{Irvine}
  \country{USA}
}
\email{iftekha@uci.edu}



\begin{abstract}
Commit messages are crucial in software development, supporting maintenance tasks and communication among developers. While Large Language Models (LLMs) have advanced Commit Message Generation (CMG) using various software contexts, some contexts developers consider to write high-quality commit messages are often missed by CMG techniques and can’t be easily retrieved or even retrieved at all by automated tools. To address this, we propose Commit Message Optimization (CMO), which enhances human-written messages by leveraging LLMs and search-based optimization. CMO starts with human-written messages and iteratively improves them by integrating key contexts and feedback from external evaluators. Our extensive evaluation shows CMO generates commit messages that are significantly more Rational, Comprehensive, and Expressive while outperforming state-of-the-art CMG methods and human messages 40.3\%-78.4\% of the time. Moreover, CMO can support existing CMG techniques to further improve message quality and generate high-quality messages when the human-written ones are left blank.
\end{abstract}




\maketitle

\section{Introduction}
\label{sec:intro}

Commit messages are essential in documenting code changes by explaining what was modified and why, as well as providing necessary context for software maintenance. High-quality messages succinctly capture the change details (``What''), the rationale (``Why''), and any additional relevant information, which is vital for assessing the commit's impact on the codebase ~\cite{tian2022makes, li2024only}. Well-written messages also help reduce software defect proneness~\cite{li2023commit}. However, developers often neglect to write high-quality messages due to time constraints or lack of motivation. This leads to vague or incomplete messages—even in projects under the Apache Software Foundation~\cite{tian2022makes, li2023commit}. This issue is widespread; for instance, Dyer et al. ~\cite{dyer2013boa} found that approximately 14\% of commit messages in over 23,000 SourceForge projects were entirely blank.

To assist developers in writing commit messages and elevate the commit message quality issue, numerous automated Commit Message Generation (CMG) techniques have been proposed by the Software Engineering (SE) research community. These methods, given a set of code changes, aim to generate descriptive commit messages. These techniques can be broadly categorized into \textit{Template-based} approaches where predefined patterns are used for generating messages \cite{linares2015changescribe, shen2016automatic}. \textit{Retrieval-based} methods which find similar past commit messages and adapt them \cite{huang2020learning, shi2022race}. \textit{Translation-based} techniques treat commit message generation as a language translation problem, translating code changes into natural language \cite{xu2019commit,liu2020atom, nie2021coregen}. \textit{Hybrid} approaches combine elements from the above techniques to enhance performance \cite{liu2020atom, shi2022race,he2023come}. However, researchers have noted that these methods often produce messages that lack key information, such as the ``What'' or ``Why,'' provide insufficient context, follow simple patterns, or are semantically unrelated to the human-written messages and difficult to read \cite{jiang2017automatically,wang2021quality,liu2018neural,dong2023revisiting}. This underscores the need for further improvements in CMG techniques to generate higher-quality messages.

Recent advancements in Large Language Models (LLMs) have significantly enhanced various software engineering tasks, including CMG. Li et al. ~\cite{li2024only} introduced a state-of-the-art approach leveraging GPT-4 with the ReAct prompting technique ~\cite{yao2022react}. Their method, Omniscient Message Generator (OMG), outperformed previous CMG techniques by incorporating a broader range of contextual information. While traditional CMG methods rely solely on code changes ~\cite{liu2020atom, shi2022race, dong2022fira}, OMG integrates additional software contexts, such as pull request titles and method/class summaries. However, since commit messages serve as one of the primary communication channels for developers, facilitating discussions on diverse software maintenance activities that often depend on various software contexts \cite{mannan2020relationship,tian2022makes,li2023commit}, it is plausible that OMG misses certain nuanced contextual information considered by developers to compose high-quality messages. This raises the question:

\textbf{RQ1:What software context information do developers consider when writing high-quality commit messages that existing CMG techniques fail to capture?}



The variety of contextual information developers consider while writing commit messages makes it difficult for automated tools to capture all the necessary human-considered contexts for generating high-quality commit messages. To address this, integrating human guidance is essential for producing commit messages that reflect these diverse contexts. Human guidance has already proven effective in tasks like code repair and assertion generation \cite{bohme2020human,zamprogno2022dynamic}, and similar methods could be beneficial in CMG. Eliseeva et al. \cite{eliseeva2023commit} introduced Commit Message Completion (CMC) by using human-typed prefixes to help LLMs generate more accurate commit messages. However, their study did not evaluate whether the generated messages included essential information like the ``What''/``Why'' information. Furthermore, they only considered the preceding commit to assist in message personalization, overlooking other critical software contexts that can impact commit message quality \cite{tian2022makes, li2023commit, li2024only} including class/method containing the changes and related issue report/pull request.

Inspired by CMC, we propose a new method called \textit{Commit Message Optimization} (CMO) to enhance the quality of commit messages by leveraging a search-based optimization approach. Starting with a human-written message, CMO iteratively refines and improves the message to optimize its quality. This method addresses the fact that human-written commit messages can often contain quality issues, such as missing essential information or poor readability \cite{tian2022makes, li2023commit}. CMO brings two key advantages; firstly, by starting from a human-written message, the model gains access to the nuanced context a developer would naturally consider for the associated code changes. This includes information that might be impractical for automated systems to retrieve but is readily available in a developer's mind. Second, using the power of LLMs, CMO can ``fix'' or optimize the commit message by incorporating widely considered contexts and addressing quality issues that the human developer may miss. The iterative process allows the LLM to enhance the message with more relevant information, improving its clarity and completeness. 

Since developers often leave commit messages empty due to time and motivation constraints \cite{dyer2013boa}, CMO is designed to address this issue by optimizing an initial message generated through few-shot prompting \cite{brown2020language}. Additionally, we investigate whether CMO can enhance existing CMG techniques, such as OMG, to improve the quality of generated messages. While CMG techniques have been extensively studied in research, their adoption among software practitioners remains limited \cite{eliseeva2023commit, dong2023revisiting}. Our goal is to bridge this gap by making CMG more practical and effective for real-world software development: (1) developers can optimize their own commit messages, mitigating the concern of CMG generating semantically irrelevant messages from scratch. (2) CMG-generated messages can be readily optimized with guidance and frequently considered contexts, which would improve CMG's effectiveness in practice. Thus, we propose the following research questions:



\textbf{RQ2: How does CMO perform compared to the state-of-the-art CMG technique, CMC technique, and human developers?}

\textbf{RQ3: How can CMO support existing CMG techniques and generate messages when the human-written messages were left blank?}

Finally, we conduct an ablation study to analyze the effectiveness of the major components of CMO in generating high-quality commit messages:

\textbf{RQ4: How do software context collection tools and search-based optimization contribute to the overall effectiveness?}

To sum up, our contributions are as follows:

\noindent1. We systematically identify \textbf{software contexts that are widely considered by developers to write high-quality commit messages but missed by existing CMG approaches}.

\noindent2. We propose a novel framework, \textbf{CMO, that optimizes existing human-written commit messages} using LLMs, software contexts, and automated evaluators.

\noindent3. Our human evaluation shows that our approach with appropriate hyper-parameters can \textbf{significantly improve human-written commit message quality and outperform existing CMG and CMC techniques}.

\noindent4. Our approach can \textbf{enhance the performance of the state-of-the-art CMG technique and generate high-quality commit messages when the human-written ones were left blank}.

\section{Background \& Related Work}
\label{sec:rw}

\subsection{Commit Message Quality}
Researchers have traditionally assessed commit message quality using syntactic features like message length, word frequency, punctuation, and imperative verb mood~\cite{chahal2018developer, chen2020project}, but these methods overlook the semantic content. To address this, Tian et al.\cite{tian2022makes} proposed that high-quality commit messages should summarize code changes (``What'') and explain their motivations (``Why''). However, they considered issue report/pull request links as providing the ``Why'' without evaluating the content of these links. Li et al.~\cite{li2023commit} addressed this by considering both commit messages and their link contents. In a separate study, Li et al.~\cite{li2024only} expanded the criteria to include additional quality expectations from developers. Our study aims to improve both CMG-generated and human-written commit messages by optimizing them based on these quality factors.

\subsection{Commit Message Generation}
Researchers have developed various CMG techniques that automate message generation from code changes. These techniques have a variety of underlying mechanisms. Template-based approaches~\cite{linares2015changescribe, shen2016automatic} use predefined templates for specific code changes but struggle with generalization and often omit the crucial ``Why'' information. Retrieval-based methods~\cite{huang2020learning, shi2022race} retrieve similar code changes and reuse their commit messages. However, this approach relies on human-written messages, which are frequently flawed—about 44\% of the commit messages in Open-Source Software (OSS) projects miss critical ``What'' or ``Why'' details~\cite{tian2022makes}. Translation-based techniques~\cite{liu2020atom, nie2021coregen} use Neural Machine Translation (NMT) to ``translate'' code changes into commit messages but a majority (90\%) of the generated messages are short, simple, and inadequate~\cite{dong2023revisiting}. Hybrid approaches~\cite{shi2022race, he2023come} combine retrieval and translation methods, yet their effectiveness can be hampered by the limitations of both techniques and issues with the fusion process~\cite{he2023come}.

Despite advancements in CMG, most approaches still rely solely on the git diff as input, overlooking associated software contexts that could enhance message quality~\cite{liu2020atom, shi2022race, dong2022fira}. Only a few studies have incorporated additional contexts, such as issue states or modified ASTs, alongside code changes~\cite{liu2020atom, wang2023delving}. Recently, Eliseeva et al.\cite{eliseeva2023commit} introduced CMC that uses the prefix of human-written messages and commit history to improve performance and tries to address the issue of generating commit messages that are semantically irrelevant to human-written ones \cite{jiang2017automatically, liu2018neural,wang2021quality}. However, the authors did not assess the semantic quality of completed messages (``What''/``Why'')~\cite{tian2022makes,li2023commit,li2024only} or consider essential software contexts crucial for producing high-quality commit messages such as pull request/issue report and the changed class/method~\cite{li2024only}.


With the advent of powerful LLMs such as ChatGPT~\cite{chatgpt} and GPT-4~\cite{gpt-4}, researchers have integrated these models into CMG techniques, achieving results comparable or superior to existing approaches~\cite{li2024only,wu2024commit,wu2025empirical}. To meet software developers' expectations for high-quality commit messages while considering broader software contexts, Li et al. \cite{li2024only} introduced OMG, which uses ReAct prompting \cite{yao2022react} with GPT-4. This approach retrieves relevant software contexts, such as issue reports, pull requests, and method/class changes, to ground commit messages in factual software repository data. Human evaluations showed that OMG had set a new state-of-the-art (SOTA) CMG, with some messages deemed better than those that humans wrote. 


However, OMG still missed several software context details including certain related code changes in commit history and project requirement, which humans commonly consider when writing high-quality commit messages. To overcome the limitations of the SOTA technique, our study introduces additional retrieval tools, leverages human-written commit messages as a ``prefix'' when available, and integrates guidance from external evaluators. These enhancements ensure higher message quality by preserving critical information and preventing omissions.

\subsection{Large Language Models \& Search-based Optimization}

In recent years, LLMs have been increasingly applied to optimization, search-based planning, and problem-solving across various domains~\cite{liu2023large, yao2024tree, yang2024large}. Optimization typically begins with an initial solution, which is iteratively refined or replaced to maximize an objective function. Traditionally, this process requires carefully formulating the optimization problem and using an external solver to compute updates. However, recent studies have shown that embedding the optimization problem in a prompt and instructing LLMs to refine solutions based on previous outputs and problem descriptions iteratively can yield strong performance \cite{yao2024tree, yang2024large}. Despite this progress, existing CMG techniques have not explored this approach's applicability. This paper addresses this gap by leveraging a state-of-the-art LLM as an optimizer to enhance commit message generation. Our approach iteratively refines commit messages by incorporating new context at each step, producing progressively improved candidates. By integrating feedback from multiple commit message quality evaluators, our method enables the LLM to generate higher-quality commit messages through a continuous optimization process.

\section{RQ1: Categorization of Missing Software Context Information}
\label{sec:rq1}



Li et al. \cite{li2024only} used GPT-4 in their OMG approach to generate commit messages, achieving state-of-the-art results in CMG. However, since commit messages are a key communication channel for discussing diverse maintenance activities across various software contexts \cite{tian2022makes, li2023commit}, we argue that developers have considerable freedom to consider any information sources that originate from various places in the software repository or even interpersonal discussions and hence OMG may miss certain crucial information in its generated messages. 

For example, as shown in Figure \ref{figure:im}, understanding the definition and usage of the invoked method \textit{getComputer} is essential to generate a commit message that semantically aligns with the human-written one. OMG generated a message as \textit{``Fix: Exclude ``master'' from testGetComputerView() test. In the testGetComputerView() method in the ComputerClientLiveTest.java file, a condition has been added to exclude the ``master'' from the test. This change ensures that the test only validates the display name of each computer in the view, excluding the ``master''.''}, which reflects its lack of knowledge or consideration of this invoked method. Instead, it only captures the textual content of the source code at a surface level (``\textit{...exclude the ``master'' from the test.}'', ``\textit{only validates the display name of each computer in the view, excluding the ``master''.}''), missing the essential reason behind the change that \textit{``master is not accessible via getComputer.''} In this study, we investigate the software contexts humans consider when writing high-quality messages that OMG failed to capture. Incorporating them in the CMG techniques may further improve the generated message quality. We present the methodology in (Section \ref{sec:rq1_method}) and the results (Section \ref{sec:res_rq1}). 

\subsection{Methodology to Answer RQ1}
\label{sec:rq1_method}

Li et al. \cite{li2024only} constructed the \textit{OMG dataset} (Section \ref{sec:dataset}), consisting of 381 commits from 32 Apache projects. Each commit in this dataset includes both a human-written commit message and an OMG-generated message. To comprehensively evaluate commit message quality, Li et al. proposed four key metrics (Section \ref{sec:hm_metrics}), scoring all messages on a 5-point Likert scale (0: poor to 4: excellent). The full scoring process is detailed in \cite{li2024only}. To ensure the use of high-quality human-written messages, we filtered out commits where the human-written message scored below 3 in any of the four metrics. This selection process resulted in 262 commits, retaining both their human-written and OMG-generated messages.

We qualitatively analyzed the code changes, high-quality human-written commit messages, and OMG-generated messages for the 262 selected commits to identify the software contexts likely considered when composing these messages. This analysis followed the open coding protocol \cite{glaser2016open}. Two researchers with more than five years of Java development independently examined the code changes and associated commit messages, coding the software contexts referenced in the messages. These coded contexts were then verified through close examination to ensure their relevance. For example, as illustrated in Figure \ref{figure:im}, if the human-written message referenced details about the invoked method \textit{getComputer}, we coded this as \textit{Excluded Callee Knowledge} (Section \ref{sec:res_rq1}). To verify its importance, we manually inspected the method's body to determine whether awareness of its functionality would support message composition. Similarly, we identified and verified contexts that contributed to the OMG-generated messages. Each newly identified context was compared against existing codes during coding to determine whether it represented a distinct category or a subset of an existing one. This iterative comparison was conducted throughout the coding process and refined through negotiated agreement \cite{forman2007qualitative}, where researchers discussed the rationale for applying specific codes and reached a consensus. Ultimately, we identified the contexts considered by human authors but overlooked by OMG. Section \ref{sec:res_rq1} details the finalized set of context codes.

\subsection{Answer to RQ1: Software Context Missed by CMG}
\label{sec:res_rq1}

\subsubsection{Context Themes and Frequency}
\label{sec:themes}
We identified seven categories of software context that humans considered while writing high-quality messages, but OMG missed. Specifically, 67.6\% (177/262) of OMG-generated messages failed to include some software contexts recognized by humans. Notably, a single commit can exhibit multiple contexts missed by OMG but considered by humans. The categories are described below:

\noindent\textbf{1. Unreferenced Software Maintenance Goals} (55.9\%, 99/177): 
This category involves developers describing code changes with reference to a specific maintenance goal, such as functional correction, new feature addition, or non-functional improvement, without citing an issue report or pull request. Instead, the commit message itself conveys the goal. For example, a message like \textit{``change required after plexus update"} \cite{adhoc} indicates that the changes ensure compatibility with an updated third-party dependency, with the maintenance goal serving as context for explaining the code changes.

Moreover, some code changes are made to resolve some personal mistakes developers have made in the past. These mistakes can include copy-paste errors, accidents due to negligence, and typos. 
For example, \textit{``fix typo: wrong if guard variable''} \cite{mistake} shows the reason for the \textit{wrong if guard variable} is the fact that some developer made a typo. In addition, some developers may simply write commit messages as \textit{``Remove author tag. Thanks Sylvain for pointing at this, this happens when you copy paste and don't think about what you're doing.''} \cite{personalmistake1} or \textit{``Remove getFilter method inadvertantly left in''} \cite{personalmistake2}.



\begin{figure*}[h]
\label{figure:examples}
    \centering
    \begin{subfigure}[t]{0.5\textwidth}
        \centering
        \includegraphics[height=1.0in]{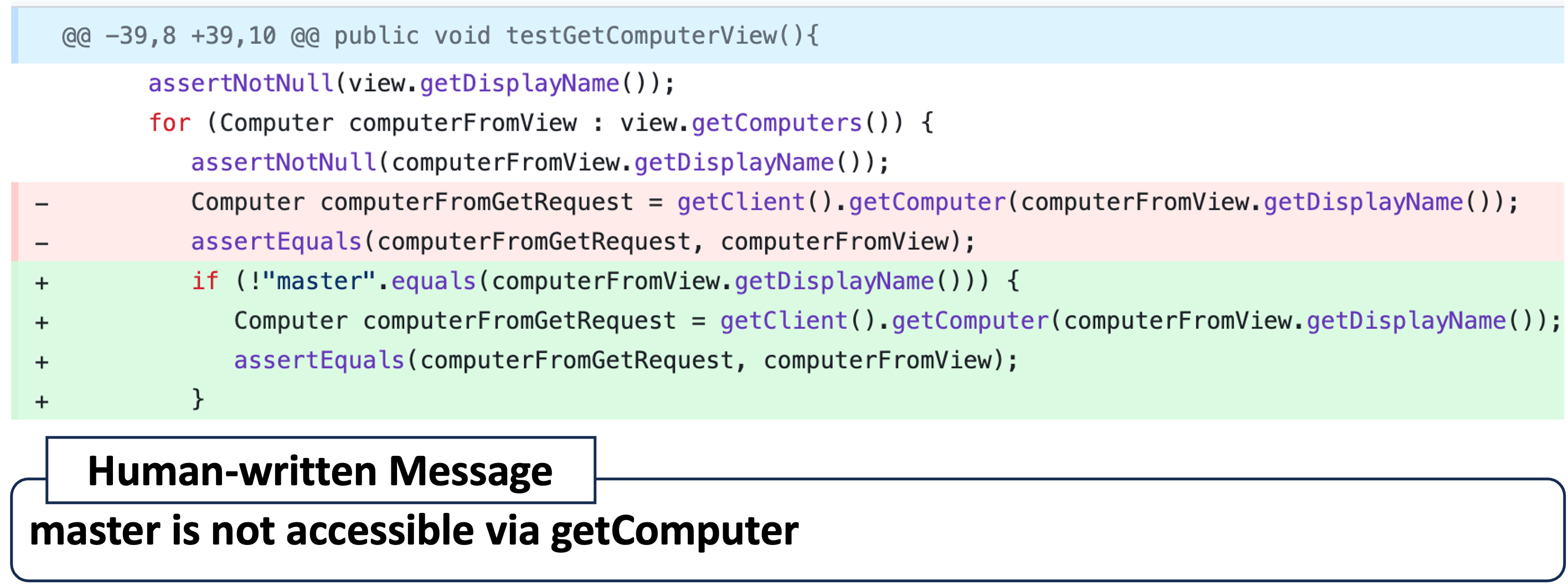}
        \caption{Example of Necessary Callee Knowledge}
        \label{figure:im}
    \end{subfigure}%
    ~ 
    \begin{subfigure}[t]{0.5\textwidth}
        \centering
        \includegraphics[height=1.0in]{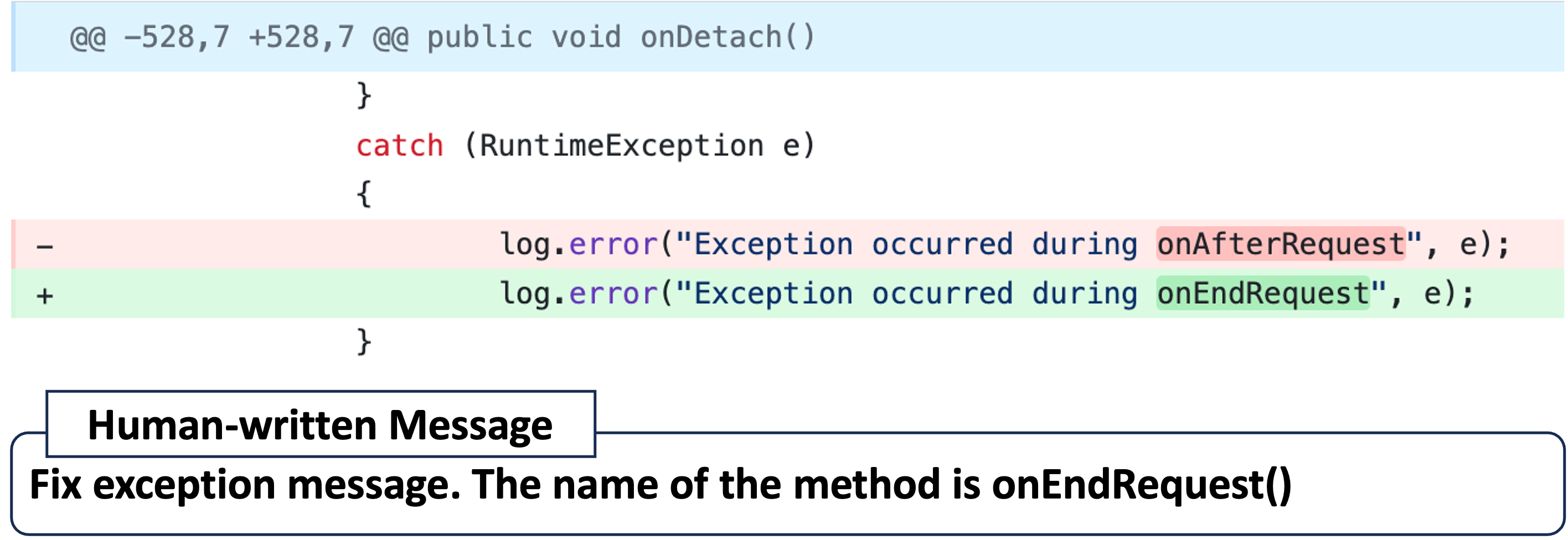}
        \caption{Example of Complete Enclosing Code Blocks}
        \label{figure:cb}
    \end{subfigure}
    \caption{Example Commits}
\end{figure*}

\noindent\textbf{2. Excluded Callee Knowledge} (24.9\%, 44/177): 
This category occurs when developers modify method calls or make changes related to method calls, leading them to describe the ``What''/``Why'' based on their knowledge of those invoked methods. However, such methods are not defined in the code change (git diff). For example, as shown in Figure \ref{figure:im} \cite{invoked}, developers used the invoked method to explain the motivation behind the changes. In this case, the addition of the \textit{if} statement is due to the fact that \textit{``master"} is inaccessible via \textit{getComputer}. Understanding the method \textit{getComputer} was essential for crafting this commit message.








\noindent\textbf{3. Implicit Project Requirements/Practices} (24.3\%, 43/177): 
Developers often implicitly rely on project requirements or recommended practices to explain code changes. Phrases like ``should,'' ``do not,'' or ``allow'' suggest that a project requirement or convention influenced the decision, though developers did not reference specific requirements directly. Instead, they use certain words to imply the necessity of the change. For example, \textit{``Don't try to config mdb destination if we aren't auto creating resources.''} \cite{projreq} suggests that the change was made to correct a violation of a project requirement, even though the specific requirement is not explicitly mentioned.

\noindent\textbf{4. Excluded Variable Data Types} (15.3\%, 27/177): Certain variables are often referenced but not defined in the code change (i.e., git diff), but either in another file or in the same file outside the code change. However, understanding the class or data types of such variables helps explain the code changes in the commit message. For example, in commit \cite{datatype}, developers added several methods, each returning a variable. The message \textit{``Add getters for private ivars''} highlights that the variables' data types and access modifiers were considered.



\noindent\textbf{5. Miscellaneous Related Code Changes in History} (9.6\%, 17/177): 
Code changes from the project's history are often referenced when writing commit messages for a given commit. These historical changes may relate to the current commit in various ways, such as fixing defect-introducing changes, enhancing earlier features, or repurposing previous modifications. For example, \textit{``Re-adding Reflection2.constructor. Removed in 671749d but used downstream in jclouds-labs''} \cite{commithist}. Here, the previous changes serve as context for composing the message.




\noindent\textbf{6. Complete Enclosing Code Blocks} (7.3\%, 13/177): The surrounding source code is often considered when writing commit messages. However, the limited length of the git diff—which represents the code changes -- may not provide sufficient context for developers or CMG approaches to fully explain or summarize the changes. Additional context, such as the entire enclosing statement block, is often needed but may not be included in the git diff. OMG missed this information because its summaries of the enclosing method or class captured only high-level functionality, leaving out important detailed source code information. Hence this theme focuses on the source code near the changed lines, as opposed to the broader summaries of the method or class considered by OMG. For example, accurately describing the changes in Figure \ref{figure:cb} \cite{codeblock} (i.e., \textit{onEndRequest}) requires access to the full enclosing \textit{try-catch} block to identify the executed method in the \textit{try} section. Relying solely on the git diff would miss this crucial context, as reflected in the associated human-written commit message.

\noindent\textbf{7. Requisite Compile/run-time Information} (5.1\%, 9/177): When developers commit code changes to fix compilation or run-time errors, they often explicitly specify this in the commit message. For example, \textit{``Resolve trivial compilation error after previous merge''} \cite{compile} indicating the changes address a compile-time error, reflecting the developers' understanding of the software’s compile-time behavior.

\section{RQ2: Effectiveness of Commit Message Optimization (CMO)}
\label{sec:cmo}

\subsection{Automated Software Context Retrieval}
\label{sec:tools}

To support the LLMs' reasoning and generation of commit messages that are more factual and grounded on the code base, we implement tools to automatically retrieve the identified software contexts in Section \ref{sec:res_rq1} that were missed by OMG. These tools aim to capture the human-considered contexts, further enhancing the quality of LLM-generated commit messages.

To extract \textit{Excluded Callee Knowledge}, \textit{Excluded Variable Data Types}, and \textit{Complete Enclosing Code Blocks} (Section \ref{sec:res_rq1}), we leveraged JavaParser \cite{javaparser}, a widely used tool for parsing Java program elements \cite{wang2020empirical,li2024only}. This allowed us to retrieve relevant information about invoked methods, variables, and statement blocks related to the modified code. Specifically, we identified the data types and class information of variables referenced in the code change (\textit{Excluded Variable Data Types}). Additionally, we extracted the smallest enclosing statement blocks, which often provide sufficient context for composing human-written commit messages (\textit{Complete Enclosing Code Blocks}). These blocks, delimited by curly brackets, included all statement types supported by JavaParser. For \textit{Excluded Callee Knowledge}, we extracted the method bodies of functions invoked within the modified code and summarized them using a state-of-the-art method-level code summarization technique \cite{geng2024large}, also employed by OMG. This approach helps manage the limited input context length of LLMs \cite{vaswani2017attention}. Our extraction process focused on retrieving invoked methods and variable data types defined within the project's source files or as part of the Java language. Identifying methods and data types from third-party libraries remains an open area for future work.

However, we did not implement automated tools to retrieve information for all software context themes in Section \ref{sec:res_rq1} due to the impracticality of automatically retrieving them and the significant noise brought by this process. To support our decision and assess the feasibility of implementing the retrieval tools for these themes, we also consulted with two senior Java software engineers with more than ten years of Java software development and commit message writing experience.

For \textit{Unreferenced Software Maintenance Goals}, code changes often address specific maintenance objectives documented in artifacts like issue reports or pull requests. However, identifying these goals without explicit references is challenging. While issue-commit link recovery techniques could help, their limited accuracy (0.1–0.5 \cite{zhang2023ealink}) makes their integration risky, potentially reducing commit message reliability. As improving these techniques is beyond our scope, we excluded them. Moreover, some goals, like personal mistakes recognized during coding, exist only in the developer’s mind and cannot be automatically retrieved.

Another approach is to compile and run the code before and after changes to detect performance improvements or bug fixes. However, this requires analyzers, tests, and environment setup. After resetting repositories to pre- and post-change versions and resolving dependencies, only 33\% compiled and ran successfully, likely due to bugs, dependency issues, or other factors. Hence, to avoid introducing significant noise, we did not develop tools for retrieving compile/run-time information to help LLMs identify maintenance goals. Similarly, we also didn't implement retrieval tools for \textit{Requisite Compile/run-time Information}.


For \textit{Implicit Project Requirements/Practices}, detailed information about the specific requirements guiding code changes is often unavailable. Creating traceability links between source code and requirements requires substantial manual effort and is typically incomplete in most projects \cite{hey2021improving}. Moreover, existing traceability recovery techniques perform poorly, with F-1 scores below 0.5 \cite{zhao2017improved,chen2019enhancing,moran2020improving,hey2021improving}, risking significant noise if applied. Since improving these methods is beyond our scope, we did not develop tools to extract linked requirements from code changes.

For \textit{Miscellaneous Related Code Changes in History}, automatically retrieving relevant historical commits is challenging. Our analysis of 17 commits where developers referenced past changes revealed four types of relevant history: (1) the most recent commit changing the same lines (23.5\%), (2) the most recent commit changing the same functions (17.6\%), (3) a specific commit affecting the same lines or functions (23.5\%), and (4) a commit impacting related code units or motivating the current change (35.4\%). Since which type of historical change is relevant depends heavily on context, any retrieval tool based on a single pattern would likely introduce significant noise. Therefore, we did not implement such a tool.

\subsection{Commit Message Optimization (CMO)}

\subsubsection{Motivation}
Since many of the contexts frequently considered by developers for writing high-quality commit messages are missed by CMG techniques and cannot be reliably extracted using automated tools (as discussed in Section \ref{sec:tools}), human-written commit messages offer a valuable alternative. These messages inherently capture software contexts that are difficult or impractical to extract automatically. Incorporating human-written messages as a reference in CMG could mitigate the need for extensive context retrieval while improving message quality. However, despite the proven success of human guidance in automating other SE tasks—such as program repair and assertion generation~\cite{bohme2020human,zamprogno2022dynamic}—its potential in CMG has been largely overlooked. In this study, we introduce \textit{Commit Message Optimization} (CMO), a technique that enhances commit message quality by optimizing human-written messages with additional software contexts frequently considered by developers. 
The following sections outline the design and implementation of CMO in detail.

\subsubsection{Objective Function}
\label{sec:obj}
In this study, we aim to enhance the quality of generated commit messages using four key metrics detailed in Section \ref{sec:hm_metrics}. To achieve this, we employed two complementary evaluators. The first, called the \textit{LLM-based Quality Evaluator}, leverages LLMs' reasoning capabilities to assess commit messages based on corresponding git diffs, focusing on how well the message reflects the code changes. The second, termed the \textit{Retrieval-based Quality Evaluator}, focuses on identifying essential software contexts—beyond just the diff--that are typically present in high-quality human-written messages. Together, these evaluators offer a holistic view of message quality. Below, we describe each evaluator in detail and explain how their outputs are combined into a unified evaluation score.




We detail the evaluators and the combined evaluation score calculation mechanism below.

\noindent\textbf{1) LLM-based Quality Evaluator}: 
We fine-tuned GPT-3.5-Turbo \cite{gpt35} (GPT-4 was unavailable for fine-tuning at the time) to automatically assess commit messages based on the four metrics in Section \ref{sec:hm_metrics} since LLMs can serve as evaluators \cite{kim2023prometheus,yao2024tree,gao2024llm}. Using the \textit{training split} of the \textit{OMG dataset}, we trained the model and evaluated on the \textit{validation split} (Section \ref{sec:dataset}). Each commit in the dataset is associated with three different messages (human-written, FIRA-generated (FIRA \cite{dong2022fira}, a CMG outperformed by OMG \cite{li2024only}), and OMG-generated), resulting in 915 training and 228 validation messages, each labeled with four human-assigned scores. We framed this as a multi-class classification task, where Likert scores (0–4) served as class labels and fine-tuned GPT-3.5-Turbo to predict scores based on git diffs and commit messages. We fine-tuned separate models to improve performance, one for each metric.
To mitigate class imbalance (e.g., only 8.6\% of messages received a score of 3 for \textit{Rationality}), we applied random oversampling \cite{kotsiantis2006handling} using \textit{imblearn} \cite{imblearn}. Table \ref{tab:llmeval} presents the classifier performance on the \textit{validation split} for each metric.

\begin{table}[h]
\caption{The Performance of LLM-based Quality Evaluators}
\label{tab:llmeval}
\resizebox{\columnwidth}{!}{%
\begin{tabular}{l|cccc}
\hline
                  & \textbf{Accuracy} & \textbf{Precision} & \textbf{Recall} & \textbf{F1} \\ \hline
Rationality       & 0.719             & 0.653              & 0.719           & 0.684       \\ \hline
Comprehensiveness & 0.719             & 0.695              & 0.719           & 0.660       \\ \hline
Conciseness       & 0.895             & 0.801              & 0.895           & 0.845       \\ \hline
Expressiveness    & 0.912             & 0.849              & 0.912           & 0.875       \\ \hline
\end{tabular}
}
\end{table}

\begin{figure}[h]
    \centering
    \centerline{\includegraphics[width=0.7\linewidth]{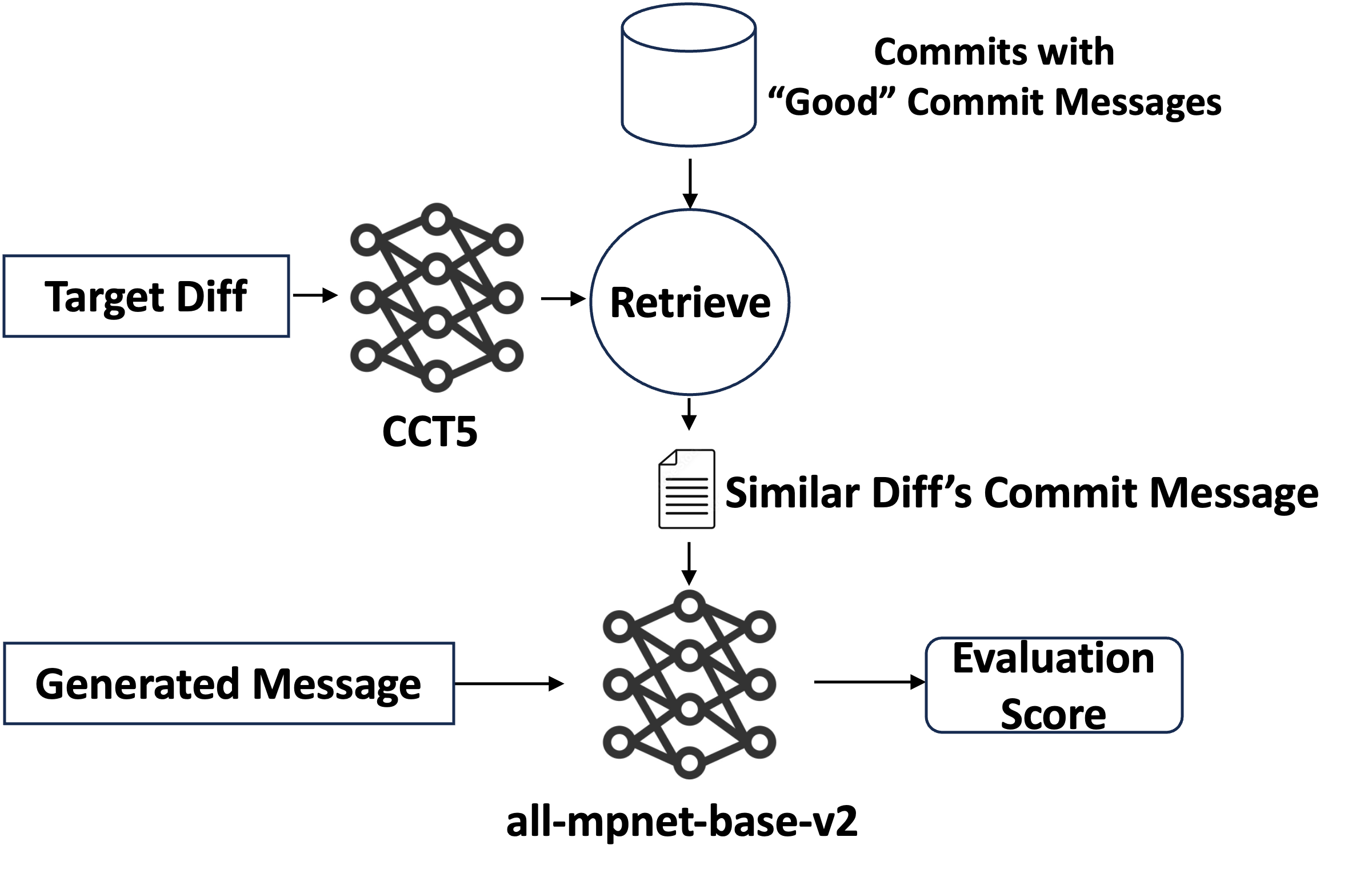}}
    \caption{Overview of Retrieval-based Quality Evaluator}
    \label{fig:retrieval-eval}
\end{figure}

\noindent\textbf{2) Retrieval-based Quality Evaluator}: 
To provide additional human guidance regarding the contexts to be considered in the objective function of optimizing the messages, especially when the initial human-written message is of low quality without valuable contexts, we introduce the \textit{Retrieval-based Quality Evaluator}. This evaluator gauges the quality of generated commit messages by measuring their semantic similarity to high-quality human-written messages that cover both the ``What'' and ``Why'' aspects, providing quality estimation by comparing with what a skilled human developer would write. As depicted in Figure \ref{fig:retrieval-eval}, the process starts by retrieving git diffs from a data corpus that are semantically similar to the target diff being optimized. The evaluator then compares the generated commit message with a high-quality human-written message, using semantic similarity as the evaluation score. In addition, the retrieved high-quality human-written messages and their git diffs are provided to the optimization process as extra guidance.  %



To construct a high-quality data corpus for retrieving similar git diffs, we collected commits from 32 Apache projects previously analyzed by OMG \cite{li2024only} and other studies \cite{mannan2020relationship,li2023commit}. To ensure the inclusion of only well-formed commit messages, we filtered commits based on whether they contained both ``What'' and ``Why'' information, using the criteria for ``Good'' commit messages defined by Tian et al.\cite{tian2022makes}. This classification was performed using the state-of-the-art model from Li et al. \cite{li2023commit}, which automatically detects the presence of ``What'' and ``Why'' components in commit messages.



The evaluator starts by identifying diffs that are most similar to the target diff. For this purpose, we employed CCT5 \cite{lin2023cct5}, a pre-trained language model designed to capture the semantic essence of code changes. We represented each git diff using the vectorized embedding of the special token [CLS] from CCT5's final encoder layer and computed cosine similarity to identify the most similar diffs to the target diff.

Next, to assess the semantic similarity between commit messages, we utilized \textit{all-mpnet-base-v2} \cite{mpnet}, a high-performing natural language model from the Sentence Transformer Leaderboard \cite{sentencetransformer}. We vectorized commit messages and measured cosine similarity between the generated message and human-written messages from the most semantically similar git diffs identified using CCT5~\cite{liu2018neural,gao2023makes,geng2024large}. This cosine similarity score served as an evaluation metric, reflecting how closely the generated message aligned with high-quality human-written messages.

To mitigate potential bias from relying on a single most similar git diff, we experimented with different retrieval sizes. Specifically, we retrieved the top 1, 5, 10, and 20 most similar git diffs and computed the final evaluation score as the average cosine similarity between the generated message and human-written messages.

\noindent\textbf{3) Combined Evaluation Score}: 
Finally, we combine the results from both evaluators to produce a single score for each of the four metrics, which serves as the objective function. To normalize the evaluation score of the \textit{Retrieval-based Quality Evaluator} (\textit{``Sim Score''}), which ranges from 0 to 1, we scaled it to a range of 0 to 4 by multiplying by 4. This allowed us to directly compare it with the \textit{LLM-based Quality Evaluator} score (\textit{``LLM Score''}). 




Additionally, we propose that the combined evaluation score can better approximate human judgment by incorporating the correlation between \textit{``Sim Score''} and human-labeled scores, as well as \textit{``LLM Score''} and human-labeled scores, into the weighting strategy. Following prior research that uses Pearson correlation to assess the effectiveness of automated metrics compared to human judgment \cite{wang2023chatgpt,liu2023g}, we conducted a Pearson correlation analysis on the scores from the \textit{validation split} of the \textit{OMG dataset}. The resulting correlation coefficients were used as weights for \textit{``Sim Score''} and \textit{``LLM Score''} in the combined score (Equation \ref{eqn:corl-eval}). Here, \textit{Sim Coeff} represents the correlation coefficient between \textit{``Sim Score''} and human-labeled scores, while \textit{LLM Coeff} represents the correlation for \textit{``LLM Score''} and human-labeled scores.

For the \textit{Retrieval-based Quality Evaluator}, we configured it to retrieve the top 10 most similar git diffs, as this setting yielded the highest correlation coefficient. Notably, since \textit{``Sim Score''} did not exhibit a significant correlation (\textit{p} > 0.05) with human-labeled scores for \textit{Conciseness}, we relied solely on \textit{``LLM Score''} for that metric.

\begin{equation}
\label{eqn:corl-eval}
\resizebox{0.9\hsize}{!}{$Metric \: Score = Sim \: Score\times 4 \times (\frac{Sim \: Coeff}{Sim \: Coeff + LLM \: Coeff}) + LLM \: Score \times (\frac{LLM \: Coeff}{Sim \: Coeff + LLM \: Coeff}) $}
\end{equation}





\subsubsection{Search-based Optimization Algorithm Design}
\label{sec:opt}
Algorithm~\ref{alg:cap} outlines the search-based optimization process.
We used the tools described in Section \ref{sec:tools} to retrieve relevant software contexts. Additionally, we incorporated all tools from OMG. A comprehensive list of all automatically retrieved contexts is provided in Table \ref{tab:contexts}.



Since our optimization starts from a human-written commit message, we enhanced OMG’s commit type classifier by including both the git diff and the human-written message in the classifier prompt. Following OMG, we evaluated the classifier on the dataset from Levin et al. \cite{levin2017boosting}, using accuracy as the evaluation metric, consistent with prior reports. With the added human-written messages, our classifier achieved 81.0\% accuracy, outperforming OMG’s classifier (53.6\%) and Levin et al.’s results (76.7\%). Since commit type information is an essential part of the message format expected by developers \cite{li2024only}, we provide it directly rather than using it as an optimization context. As a result, we consider seven software contexts in total for commit message optimization.

\begin{table}[h]
\caption{Automatically Retrieved Software Contexts}
\label{tab:contexts}
\resizebox{\columnwidth}{!}{%
\begin{tabular}{c|c}
\hline
Important File Information & Commit Type Information      \\ \hline
Pull Request/Issue Report  & Method Body Summary           \\ \hline
Class Body Summary        & Complete Enclosing Code Blocks  \\ \hline
Excluded Callee Knowledge  & Excluded Variable Data Types              \\ \hline
\end{tabular}
}
\end{table}


To simplify our search-based algorithm for optimizing a single objective function, we summed the four metrics from the \textit{Combined Evaluation Score} (Section \ref{sec:obj}, referred to as the \textit{optimization score}, which is returned by the \textit{EVALUATE} function.) 
The algorithm first updates the human-written commit message using each available context individually (Table \ref{tab:contexts}), generating different commit message candidates (\textit{msg\_candidate}). At each subsequent step, the candidate with the highest optimization score is dequeued from the \textit{priority queue} and updated with the contexts that haven't been considered yet, generating further candidates for improvement (lines 11-17).


We implemented the \textit{UPDATE} function by prompting GPT-4, as it has demonstrated SOTA performance in CMG through prompting \cite{li2024only}. The prompt included the target git diff, a definition of git diff, the expected commit message format\cite{li2024only}, and explanations of the four evaluation metrics and their scoring criteria \cite{li2024only}. To help GPT-4 optimize for higher scores from the \textit{Retrieval-based Quality Evaluator} (which contributes to the overall \textit{optimization score}), we also provided the top 10 git diffs most similar to the target diff, along with their corresponding commit messages. GPT-4 was explicitly instructed to improve the existing commit message—whether human-written or a previous candidate—rather than generating a new message from scratch.

We incorporated multiple stopping criteria alongside a fixed \texttt{step\_limit} to control the optimization process. First, we introduced a dynamic score improvement threshold (\texttt{improve\_threshold}) that decreases as the number of optimization steps grows (line 7), based on the assumption that improvement naturally diminishes as candidate messages approach higher quality. Initially, the threshold is set as a percentage (\texttt{p}) of the \textit{optimization score} of the human-written message (line 3), with a minimum threshold (\texttt{min\_threshold}) to prevent it from nearing zero (lines 8–9). The optimization halts if the score improvement between the latest \texttt{highest\_score} and the one updated two steps earlier is less than the threshold (lines 22–23), enabling longer optimization for low-quality messages and early stopping for high-quality ones. Additionally, we set GPT-4's temperature to zero to generate deterministic outputs \cite{peng2023towards,li2024only}. However, if the improvement across two steps falls below the threshold, we increase the temperature to allow GPT-4 to produce more diverse candidates, potentially surpassing the threshold and continuing the optimization process \cite{liu2023large}.

\begin{algorithm}[hbt!]
\caption{Search-based Optimization}\label{alg:cap}
\begin{algorithmic}[1]
\Require{git diff (diff), available contexts, human msg}
\Ensure{optimized msg, evaluation score}
\State $step \gets 0$, $step\_limit \gets N$
\State $highest\_score \gets $ \texttt{EVALUATE(diff, human msg)}
\State $improve\_threshold \gets highest\_score \times p$, $min\_threshold \gets \frac{improve\_threshold}{step\_limit}$
\State $priority\_queue$\texttt{.enqueue(human msg)}
\While{$step \textless step\_limit$}
    \State $step \gets step+1$
    \State $improve\_threshold \gets \frac{(improve\_threshold \times (step\_limit - step))}{step\_limit}$
    \If{$improve\_threshold \textless min\_threshold$}
        \State $improve\_threshold \gets min\_threshold$
    \EndIf
    \State $cur\_msg \gets priority\_queue${.dequeue()}
    \For{\texttt{each context in available contexts}}
        \State $msg\_candidate \gets $ \texttt{UPDATE($cur\_msg$, diff, context, considered contexts, EVALUATE(diff, $cur\_msg$))}
        \State $priority\_queue$\texttt{.enqueue($msg\_candidate$)}
    \EndFor
    \State $priority\_queue$\texttt{.sort(key = evaluation score)}
    \State $cur\_score \gets $ \texttt{EVALUATE(diff, $priority\_queue$[0].msg)}
    \If{$cur\_score \textgreater highest\_score$}
        \State $highest\_score \gets cur\_score$
        \State optimized msg $ \gets $ \texttt{$priority\_queue$[0].msg}
        \State evaluation score $ \gets highest\_score $
        \If{$highest\_score - highest\_score_{updated\_two\_iterations\_ago}$ $\textless improve\_threshold$}
            \State \texttt{break}
        \EndIf
    \EndIf
\EndWhile
\end{algorithmic}
\end{algorithm}

\subsection{Methodology to Answer RQ2}

\subsubsection{Baselines}
\label{sec:baselines}
We selected OMG\cite{li2024only}, CMC \cite{eliseeva2023commit} (chosen because both CMC and our approach are guided by human input), and lastly, human-written commit messages as baselines.
\subsubsection{Commit Message Evaluation Metrics}
\label{sec:hm_metrics}
Li et al. \cite{li2024only} proposed four metrics to comprehensively evaluate the quality of commit messages. In this work, we also used these metrics: 

\noindent\textbf{Rationality}: Assesses whether the commit message provides a logical explanation for the code changes (the ``Why'' information) and clearly states the commit type. \textbf{Comprehensiveness}: Evaluates whether the commit message includes a summary of the code changes (the ``What'' information) and covers all affected files. \textbf{Conciseness}: Measures whether the commit message conveys essential information succinctly, ensuring readability and ease of understanding. \textbf{Expressiveness}: Reflects whether the commit message is grammatically correct and fluent. We also adopted traditional metrics that are widely used in previous CMG works including BLEU \cite{papineni2002bleu}, METEOR \cite{banerjee2005meteor}, and ROUGE-L \cite{lin2004automatic} where the human-written messages are used as references.







\subsubsection{Dataset}
\label{sec:dataset}

Our dataset contains 500 commits from the 32 Apache Java projects studied by OMG \cite{li2024only} consisting two sub-datasets:

\noindent\textbf{OMG dataset}: Li et al. \cite{li2024only} collected 381 commits where each commit's code diff is paired with three commit messages: one human-written, one generated by OMG, and one by FIRA \cite{dong2022fira}, a CMG technique that OMG outperformed. We randomly sampled 80\% (305) of the commits from the \textit{OMG dataset} to obtain the \textit{training split}, which was used to train the \textit{LLM-based Quality Evaluator} (Section \ref{sec:obj}). The remaining 20\% (76) were used in the \textit{validation split} to evaluate the performance of the \textit{LLM-based Quality Evaluator} and calculate the correlation in the weighting strategy of the \textit{Combined Evaluation Score}. Thus, all the components and parameters of CMO are finalized on these two splits of this sub-dataset. 


\noindent\textbf{CMO dataset}: To prevent data leakage and evaluate the effectiveness of CMO, we collected commits from the Apache projects whose commit dates were after the knowledge cutoff date of GPT-4 and GPT-3.5-Turbo (December 2023) \cite{openaimodels}. Note that we only collected commits of similar size that could be processed by OMG to ensure a fair comparison. From this pool, we randomly sampled 119 commits out of 171 eligible commits (95\% confidence level and a 5\% margin of error) due to the financial cost of using the OpenAI API to run all techniques in this study (Section \ref{sec:rq2}, \ref{sec:rq3}, \ref{sec:rq4}) and the significant manual effort required for human evaluation (Section \ref{sec:humaneval}, \ref{sec:rq3_method}, \ref{sec:rq4_method}).

\subsubsection{Human Evaluation}
\label{sec:humaneval}
We also conducted a human evaluation in which two researchers, each with over five years of Java development experience, independently evaluated 119 commits from the \textit{CMO dataset} (Section \ref{sec:dataset}). For each commit, they ranked four commit messages—comprising the human-written message optimized by CMO and baseline messages (human-written, OMG, and CMC)—based on the four evaluation metrics (Section \ref{sec:hm_metrics}). To ensure fairness, the evaluators were blinded to the sources of all messages, including the human-written ones, but had access to all relevant software contexts associated with the code changes.

As part of the human evaluation, we also surveyed 22 Apache OSS developers to assess commit messages, following university-approved Institutional Review Boards (IRB) protocols. Each developer received a survey containing 10 randomly sampled git diffs (from the 119 in the \textit{CMO dataset}), along with four corresponding messages—human-written, OMG-generated, CMC-completed, and CMO-optimized—for each diff. Developers ranked the messages based on the four evaluation metrics (Section \ref{sec:hm_metrics}). To avoid overwhelming participants and ensure higher completion rates, we limited the survey to 10 diffs instead of using the full set, as large workloads can reduce participation \cite{smith2013improving}. As in the researcher evaluation, developers were blinded to message origins but could access relevant software contexts.


\subsubsection{Hyper-parameter Tuning}
\label{sec:hypm}

As shown in Algorithm \ref{alg:cap}, several hyper-parameters that may alter the behavior of CMO exist.
To optimize CMO's performance, we explored a range of hyper-parameters: percentage value \textit{p} (5, 10, 15, 20), \textit{temperature} (0.5, 1), and \textit{step\_limit} (10, 30, 50, 60). Following a Grid Search approach \cite{jimenez2008finding}, we evaluated all possible combinations of these hyper-parameters. Due to budget constraints associated with using the OpenAI API \cite{openaiapi}, we randomly sampled 10 commits from the \textit{validation split} of the \textit{OMG dataset} and optimized their human-written messages. Four authors independently evaluated the optimized messages using the evaluation metrics described in Section \ref{sec:hm_metrics}. Based on these evaluations, we selected \textit{p} = 5, temperature = 1, and \textit{step\_limit} = 50 as the final CMO settings used in subsequent experiments.



\subsection{Answer to RQ2: Effectiveness of CMO}
\label{sec:rq2}
Table \ref{tab:researcher-eval} (white rows) presents the researchers' evaluations, showing the relative rankings where both evaluators agreed. The table indicates whether CMO-optimized messages were ranked \textit{Higher}, \textit{Equal}, or \textit{Lower} compared to the three baselines. The highest percentages for each metric are highlighted. For comparisons between CMO and human, CMO-optimized messages were considered better by both researchers for 42.9\% of the commits in terms of \textit{Rationality}, 63.0\% in \textit{Comprehensiveness}, and 69.8\% \textit{Expressiveness}. Compared with OMG, CMO-optimized messages were considered better for 42.9\% of the commits in terms of \textit{Rationality}, 40.3\% in \textit{Comprehensiveness}, and 44.0\% \textit{Expressiveness}. Table \ref{tab:researcher-eval} (grey rows) shows developers' evaluations (all 22 developers completed the survey), where each cell shows how often CMO-optimized messages were relatively ranked. These results align with the researchers' evaluation, reinforcing that CMO outperforms OMG/CMC and enhances human-written messages, though lags behind in \textit{Conciseness}.

\begin{table}[h]
\caption{Human Evaluation (by researchers (First Three Rows) and by developers (Last Three Rows)) Results on Commit Messages (RQ2)}
\label{tab:researcher-eval}
\resizebox{\columnwidth}{!}{%
\begin{tabular}{l|cccccccccccc}
\hline
                      & \multicolumn{3}{c}{\textbf{Rationality}}          & \multicolumn{3}{c}{\textbf{Comprehensiveness}}    & \multicolumn{3}{c}{\textbf{Conciseness}}           & \multicolumn{3}{c}{\textbf{Expressiveness}}       \\ \cline{2-13} 
                      & \textbf{Higher} & \textbf{Equal} & \textbf{Lower} & \textbf{Higher} & \textbf{Equal} & \textbf{Lower} & \textbf{Higher} & \textbf{Equal} & \textbf{Lower}  & \textbf{Higher} & \textbf{Equal} & \textbf{Lower} \\ \hline
\textbf{CMO VS Human} & \textbf{42.9\%} & 5.9\%          & 0.8\%          & \textbf{63.0\%} & 6.7\%          & 0.8\%          & 8.4\%           & 5.9\%          & \textbf{30.3\%} & \textbf{69.8\%} & 7.8\%          & 22.4\%         \\ \hline
\textbf{CMO VS OMG}   & \textbf{42.9\%} & 5.0\%          & 2.5\%          & \textbf{40.3\%} & 4.2\%          & 5.9\%          & 0.8\%           & 2.5\%          & \textbf{36.1\%} & \textbf{44.0\%} & 13.8\%         & 42.2\%         \\
\textbf{CMO VS CMC}   & \textbf{58.8\%} & 5.9\%          & 1.7\%          & \textbf{67.2\%} & 4.2\%          & 0.8\%          & 4.2\%           & 4.2\%          & \textbf{34.5\%} & \textbf{78.4\%} & 0.9\%          & 20.7\%         \\ \hline
\textbf{CMO VS Human} & \textbf{74.5\%} & 13.7\%         & 11.8\%         & \textbf{78.2\%} & 15.0\%         & 6.8\%          & 10.9\%          & 12.7\%         & \textbf{76.4\%} & \textbf{61.4\%} & 25.5\%         & 13.1\%         \\ \hline
\textbf{CMO VS OMG}   & \textbf{67.7\%} & 8.2\%          & 24.1\%         & \textbf{65.9\%} & 15.9\%         & 18.2\%         & 20.0\%          & 29.1\%         & \textbf{50.9\%} & \textbf{52.3\%} & 34.1\%         & 13.6\%         \\
\textbf{CMO VS CMC}   & \textbf{76.4\%} & 15.4\%         & 8.2\%          & \textbf{74.1\%} & 17.3\%         & 8.6\%          & 9.1\%           & 13.2\%         & \textbf{77.7\%} & \textbf{53.6\%} & 31.8\%         & 14.6\%         \\ \hline
\end{tabular}
}
\end{table}

We also evaluated the performance using automated evaluators by calculating the average scores for the \textit{Combined Evaluation Score} with Equation \ref{eqn:corl-eval} (Section \ref{sec:obj}), averaged across all commits in the \textit{CMO dataset}. We conducted Welch's t-test \cite{ruxton2006unequal} and Cohen's D \cite{diener2010cohen} to assess the statistical significance of score differences between approaches. As shown in Table \ref{tab:evaluators-eval}, CMO statistically significantly outperforms OMG with medium effect size (Cohen’s D larger than 0.5) for all metrics.
Additionally, CMO effectively optimizes human-written messages with large effect size (Cohen’s D larger than 0.8) for all metrics.

Aligning with the findings of Li et al. \cite{li2024only}, our analysis using traditional automatic evaluation metrics (Section \ref{sec:hm_metrics}) fails to capture the performance differences between CMO and CMC. Notably, CMC scores significantly higher in BLEU, METEOR, and ROUGE-L (e.g., BLEU for CMC is 24.67, while for CMO, it is only 7.51). Since these metrics rely on human-written messages as references, they cannot assess whether CMO has successfully optimized those messages. This highlights the limitations of traditional metrics in evaluating CMG tasks, reinforcing the conclusions of Li et al. \cite{li2024only}.



\begin{mdframed}[roundcorner=10pt]
\textbf{Finding 1: CMO effectively optimizes the human-written commit messages and outperforms state-of-the-art CMG/\\CMC techniques in \textit{Rationality}, \textit{Comprehensiveness}, and \textit{Expressiveness}.} 
\end{mdframed}

\begin{figure*}[t]
     \centering
     \includegraphics[width=0.75\textwidth]{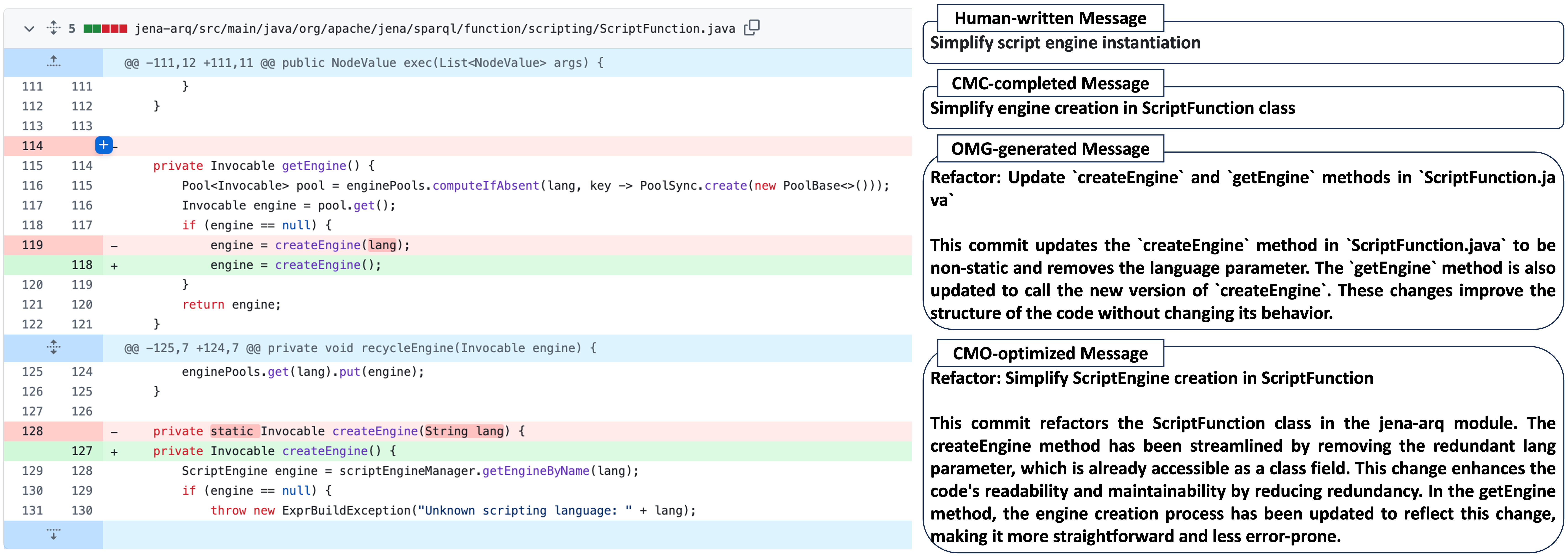}
     \caption{Example of Code Refactoring}
     \label{fig:success}
\end{figure*}

While CMO demonstrates superior performance over OMG/CMC and effectively optimizes human-written messages, its \textit{Conciseness} remains lacking. Prior research indicates that LLMs tend to generate more detailed and longer commit messages compared to humans \cite{eliseeva2023commit,zhang2024automatic, li2024only}, which may explain this outcome. Figure \ref{fig:success} \cite{successfulcase} illustrates an example where CMO optimizes a human-written commit message and outperforms both OMG and CMC, as confirmed by researchers and automated evaluators.


\begin{table}[h]
\caption{Automatic Evaluation Results (by evaluators) on Commit Messages}
\label{tab:evaluators-eval}
\resizebox{\columnwidth}{!}{%
\begin{tabular}{l|ccccc}
\hline
                    & \textbf{Rationality} & \textbf{Comprehensiveness} & \textbf{Conciseness} & \textbf{Expressiveness} & \textbf{Total}  \\ \hline
\textbf{CMO}        & \textbf{3.057}       & \textbf{3.285}             & 4                    & \textbf{3.276}          & \textbf{13.618} \\
\textbf{OMG}        & 2.880                & 3.217                      & 4                    & 3.208                   & 13.306          \\
\textbf{CMC}        & 2.617                & 3.145                      & 4                    & 3.179                   & 12.939          \\
\textbf{Human}      & 1.958                & 2.414                      & 4                    & 3.144                   & 11.516          \\ \hline
\textbf{CMO-Search} & 2.893                & 3.185                      & 4                    & 3.201                   & 13.280          \\
\textbf{CMO-File}   & 2.780                & 3.055                      & 4                    & 3.233                   & 13.067          \\
\textbf{CMO-Type}   & 3.025                & 3.278                      & 4                    & 3.268                   & 13.570          \\
\textbf{CMO-MB}     & 3.035                & 3.266                      & 4                    & \textbf{3.276}          & 13.577          \\ \hline
\textbf{CMO-OMG}    & 3.033                & 3.266                      & 4                    & 3.258                   & 13.557          \\
\textbf{CMO-blank}  & 3.045                & 3.269                      & 4                    & 3.267                   & 13.582          \\ \hline
\end{tabular}
}
\end{table}

\section{RQ3: CMO for CMG}
\label{sec:rq3}

\subsection{Methodology to Answer RQ3}
\label{sec:rq3_method}
Since human-written commit messages are sometimes left blank \cite{dyer2013boa}, we first prompted GPT-4 to generate an initial message and then applied CMO to optimize that message instead of the human-written one. The prompt included the target git diff, a definition of git diff, the expected message format\cite{li2024only}, and ten similar git diffs with their high-quality human-written commit messages. We refer to this CMO variant as \textit{CMO-blank}. Furthermore, beyond optimizing human-written messages as in RQ2 (Section \ref{sec:rq2}), we investigated whether CMO can enhance existing CMG techniques by generating higher-quality messages. To that end, we selected the state-of-the-art CMG technique OMG \cite{li2024only} and used its generated messages as initial inputs for CMO to further optimize, denoted as \textit{CMO-OMG}. We generated messages for the 119 commits in the \textit{CMO dataset} (Section \ref{sec:dataset}) and used the similar human evaluation used in RQ2 (Section \ref{sec:humaneval}).

\subsection{Answer to RQ3: Effectiveness of CMO Supporting CMG and Blank Initial Message}

Table \ref{tab:researcher-eval-cmovs} presents researchers' ranking results for \textit{CMO-OMG}, focusing on cases where both researchers agreed. Compared to baselines (top of Table \ref{tab:researcher-eval-cmovs}), \textit{CMO-OMG}-optimized messages were judged superior for 21.0\%–52.9\% of commits in \textit{Rationality}, 26.1\%–63.9\% in \textit{Comprehensiveness}, and 27.7\%–74.8\% in \textit{Expressiveness}. These findings are consistent with Table \ref{tab:evaluators-eval}, where \textit{CMO-OMG} achieves statistically significantly higher automated scores on \textit{Rationality}, \textit{Comprehensiveness}, and \textit{Expressiveness} compared to baselines.


Similarly, the middle section of Table \ref{tab:researcher-eval-cmovs} shows results for \textit{CMO-blank}, which performs comparably: outperforming baselines for 21.0\%–55.5\% of commits in \textit{Rationality}, 25.2\%–62.2\% in \textit{Comprehensiveness}, and 36.1\%–77.3\% in \textit{Expressiveness}. Table \ref{tab:evaluators-eval} further confirms these improvements with statistical significance. These results demonstrate that CMO can enhance messages produced by CMG techniques like OMG and generate high-quality messages when human-written messages are missing. However, like CMO, these variants lag behind in \textit{Conciseness}.

When comparing \textit{CMO-OMG} and \textit{CMO-blank}, both achieve similar quality in Rationality and Comprehensiveness for 25.2\%–26.1\% of commits, though \textit{CMO-OMG} has a slight edge. This suggests that CMG techniques like OMG can still benefit CMO by providing stronger initial messages, helping achieve higher final quality than starting from scratch. Thus, depending on organizational needs and resource availability, developers can choose whether to start CMO from an OMG-generated message or from empty.

\begin{table}[h]
\caption{Human Evaluation (by researchers) Results on Commit Messages (RQ3)}
\label{tab:researcher-eval-cmovs}
\resizebox{\columnwidth}{!}{%
\begin{tabular}{l|cccccccccccc}
\hline
                              & \multicolumn{3}{c}{\textbf{Rationality}}           & \multicolumn{3}{c}{\textbf{Comprehensiveness}}     & \multicolumn{3}{c}{\textbf{Conciseness}}           & \multicolumn{3}{c}{\textbf{Expressiveness}}                                        \\ \cline{2-13} 
                              & \textbf{Higher} & \textbf{Equal}  & \textbf{Lower} & \textbf{Higher} & \textbf{Equal}  & \textbf{Lower} & \textbf{Higher} & \textbf{Equal} & \textbf{Lower}  & \textbf{Higher} & \textbf{Equal}             & \textbf{Lower}                      \\ \hline
\textbf{CMO-OMG VS OMG}       & \textbf{21.0\%} & 20.2\%          & 3.4\%          & \textbf{26.1\%} & 20.2\%          & 1.7\%          & 0.8\%           & 10.1\%         & \textbf{41.2\%} & 27.7\%          & 31.9\%                     & \textbf{40.3\%}                     \\
\textbf{CMO-OMG VS CMC}       & \textbf{52.9\%} & 2.5\%           & 5.0\%          & \textbf{63.9\%} & 3.4\%           & 2.5\%          & 5.0\%           & 2.5\%          & \textbf{42.0\%} & \textbf{74.8\%} & 3.4\%                      & 21.8\%                              \\
\textbf{CMO-OMG VS Human}     & \textbf{38.7\%} & 3.4\%           & 10.1\%         & \textbf{61.4\%} & 3.4\%           & 5.0\%          & 6.7\%           & 2.5\%          & \textbf{34.5\%} & \textbf{73.1\%} & 3.4\%                      & 23.5\%                              \\ \hline
\textbf{CMO-blank VS OMG}     & \textbf{21.0\%} & 10.9\%          & 6.7\%          & \textbf{25.2\%} & 8.4\%           & 5.0\%          & 1.7\%           & 4.2\%          & \textbf{37.0\%} & 36.1\%          & \multicolumn{1}{l}{19.2\%} & \multicolumn{1}{l}{\textbf{44.5\%}} \\
\textbf{CMO-blank VS CMC}     & \textbf{55.5\%} & 2.5\%           & 3.4\%          & \textbf{62.2\%} & 4.2\%           & 4.2\%          & 3.4\%           & 2.5\%          & \textbf{39.5\%} & \textbf{77.3\%} & 2.5\%                      & 20.2\%                              \\
\textbf{CMO-blank VS Human}   & \textbf{37.0\%} & 2.5\%           & 10.1\%         & \textbf{58.0\%} & 3.4\%           & 4.2\%          & 5.9\%           & 2.5\%          & \textbf{34.5\%} & \textbf{69.7\%} & 5.0\%                      & 25.2\%                              \\ \hline
\textbf{CMO-OMG VS CMO-blank} & 15.2\%          & \textbf{26.1\%} & 8.4\%          & 16.0\%          & \textbf{25.2\%} & 8.4\%          & \textbf{16.0\%} & 8.4\%          & 10.9\%           & 36.1\%          & 26.9\%                     & \textbf{37.0\%}                     \\ \hline
\textbf{CMO VS CMO-OMG}       & \textbf{23.5\%} & 13.4\%          & 3.4\%          & \textbf{26.9\%} & 16.0\%          & 7.6\%          & \textbf{20.2\%} & 5.9\%          & 14.3\%          & \textbf{47.1\%} & 16.8\%                     & 36.1\%                              \\
\textbf{CMO VS CMO-blank}     & \textbf{25.2\%} & 16.0\%          & 3.4\%          & \textbf{23.5\%} & 21.8\%          & 3.4\%          & \textbf{21.0\%} & 5.0\%          & 10.9\%          & \textbf{50.4\%} & 19.3\%                     & 30.3\%                              \\ \hline
\end{tabular}
}
\end{table}

\begin{mdframed}[roundcorner=10pt]
\textbf{Finding 2: CMO can improve the quality of existing CMG technique generated message. CMO can also generate high-quality messages when initial human-written messages are empty.} 
\end{mdframed}


\section{RQ4: Ablation Study}
\label{sec:rq4}

\subsection{Methodology to Answer RQ4}
\label{sec:rq4_method}
The results of RQ2 and RQ3 demonstrate that CMO optimizing human-written messages outperforms other techniques. Therefore, we selected this version of CMO to conduct an ablation study, assessing the effectiveness of its two key components: the automated software context collection tools and the search-based optimization, which includes the automated evaluators. Due to financial constraints of OpenAI API usage \cite{openaiapi}, we could not ablate all seven context collection tools individually to produce seven distinct variants. Instead, we focused on the three most frequently used tools identified in Section \ref{sec:rq2}: \textit{Important File Information}, \textit{Method Body Summary}, and \textit{Excluded Variable Data Types}. This led to creating three CMO variants--\textit{CMO-File}, \textit{CMO-MethodBody}, and \textit{CMO-Type}--each omitting one tool while retaining the remaining six.

To assess the role of search-based optimization, we developed another variant, \textit{CMO-Search}, which removes the search algorithm and simply feeds all contexts retrieved by the seven tools directly into GPT-4 to optimize human-written messages.
Two researchers independently evaluated 119 commits from the \textit{CMO dataset}, ranking five messages for each commit: the human-written message optimized by the complete CMO and four messages optimized by its corresponding component-ablated variants. The rankings were based on the four established metrics (Section \ref{sec:hm_metrics}). 

\begin{table}[h]
\caption{Human Evaluation (by researchers) Results on Commit Messages (RQ4)}
\label{tab:researcher-eval-ablation}
\resizebox{\columnwidth}{!}{%
\begin{tabular}{l|cccccccccccc}
\hline
                           & \multicolumn{3}{c}{\textbf{Rationality}}           & \multicolumn{3}{c}{\textbf{Comprehensiveness}}     & \multicolumn{3}{c}{\textbf{Conciseness}}           & \multicolumn{3}{c}{\textbf{Expressiveness}}        \\ \cline{2-13} 
                           & \textbf{Higher} & \textbf{Equal}  & \textbf{Lower} & \textbf{Higher} & \textbf{Equal}  & \textbf{Lower} & \textbf{Higher} & \textbf{Equal}  & \textbf{Lower} & \textbf{Higher} & \textbf{Equal}  & \textbf{Lower} \\ \hline
\textbf{CMO VS CMO-Search} & 17.7\%          & \textbf{26.1\%} & 5.9\%         & 13.4\%          & \textbf{20.2\%} & 4.2\%          & 5.0\%          & 10.9\% & \textbf{17.6\%}         & 25.2\%          & 28.6\% & \textbf{46.2\%}         \\ \hline
\textbf{CMO VS CMO-File}   & 16.8\%          & \textbf{30.3\%} & 2.5\%          & 21.0\%          & \textbf{25.2\%} & 0.8\%          & 10.9\%          & 12.6\% & \textbf{15.1\%}         & 34.5\%          & 30.3\% & \textbf{35.3\%}          \\
\textbf{CMO VS CMO-Type}   & 7.6\%          & \textbf{39.5\%} & 0.8\%          & 9.2\%          & \textbf{37.0\%} & 0.8\%          & 32.8\%          & \textbf{42.0\%} & 25.2\%         & 28.6\%           & \textbf{37.0\%} & 34.5\%         \\
\textbf{CMO VS CMO-MB}     & 9.2\%          & \textbf{39.5\%} & 0.8\%          & 13.4\%          & \textbf{32.8\%} & 0.8\%          & 10.1\%          & \textbf{15.1\%} & 14.3\%         & 31.9\%          & \textbf{36.1\%} & 31.9\%         \\ \hline
\end{tabular}
}
\end{table}




\subsection{Answer to RQ4: Effect of Components}
As shown in Table \ref{tab:evaluators-eval}, removing any of the frequently used tools lowers CMO's scores.
However, these reductions are not statistically significant (p-value > 0.05) when comparing CMO to \textit{CMO-MethodBody} and \textit{CMO-Type}. In contrast, \textit{CMO-Search} and \textit{CMO-File} exhibit statistically significant decreases in average scores, with medium effect sizes, across all three metrics. 

Table \ref{tab:researcher-eval-ablation} presents the relative rankings where both researchers agreed, indicating whether CMO-optimized messages were ranked \textit{Higher}, \textit{Equal}, or \textit{Lower} compared to each ablated variant. When compared to \textit{CMO-Search} and \textit{CMO-File}, CMO-optimized messages were considered superior in \textit{Rationality} for 16.8\%–17.7\% of the commits and in \textit{Comprehensiveness} for 13.4\%–21.0\%. In contrast, when compared with \textit{CMO-MethodBody} and \textit{CMO-Type}, a smaller proportion of CMO-optimized messages were ranked higher—only 7.6\%–9.2\% in \textit{Rationality} and 9.2\%–13.4\% in \textit{Comprehensiveness}. These results highlight two key insights: (1) the search-based optimization component significantly improves message quality, as evidenced by CMO's advantage over \textit{CMO-Search}, and (2) the \textit{Important File Information} plays a crucial role in guiding message optimization to focus on the files that matter most to developers, as previously noted by Li et al. \cite{li2024only}.

\begin{mdframed}[roundcorner=10pt]
\textbf{Finding 3: Context collection tools and search-based optimization impact the quality of the messages.} 
\end{mdframed}

\section{Threats to Validity}
\label{sec:ttv}

In this section we list potential threats impacting the validity of our experiments. 

\textbf{Construct Validity:} 
To ensure the quality of our prompts, we followed best practices \cite{awesomeprompt,ibm} and a trial-and-error process where multiple authors manually evaluated the quality of the optimized messages on a sample of commits. 
Potential subjectivity in human evaluation was addressed by providing definitions of metrics and examples to guide the process where multiple people participated. 


\textbf{Internal Validity:} 
The \textit{OMG dataset} contains 381 commits, which may affect both the performance of the \textit{LLM-based Quality Evaluator} and the overall effectiveness of CMO. Nevertheless, this dataset is the only available set manually labeled with the evaluation metrics described in Section \ref{sec:hm_metrics}, enabling us to train the evaluator and fine-tune CMO’s parameters. To avoid data leakage, we collected 119 commits. Due to the substantial manual effort involved in human evaluation and the financial cost of using the OpenAI API, we did not analyze thousands of commits.

\textbf{External Validity:} We experimented with only the commits whose git diffs can be processed and generated messages by OMG. Our findings may not be generalizable to all OSS projects, all types of commits and to all LLMs.

\section{Conclusion and Future Work}
\label{sec:conclusion}
In this study, we identify software contexts considered by humans but missed by the state-of-the-art CMG technique, OMG. We propose \textit{Commit Message Optimization} (CMO), which leverages GPT-4 to optimize human-written messages by incorporating these missed contexts. Several automated quality evaluators serve as the objective function, aiming to maximize message quality. Our results demonstrate that CMO outperforms both OMG and CMC. Moreover, CMO can be applied to existing CMG techniques to further improve message quality, and generate high-quality messages even when initial human-written messages are empty.

Future work entails exploring how enhancing the quality of the commit messages, while preserving human-considered information, impacts other SE tasks that rely on commit message such as security patch identification \cite{zhou2021spi}, patch correctness prediction \cite{tian2022change}, code refactoring recommendation \cite{rebai2020recommending}, and defect prediction \cite{eken2019predicting}.

\bibliographystyle{ACM-Reference-Format}
\bibliography{fse}

\end{document}